\documentclass[aip,jmp,amsmath,amssymb,reprint]{revtex4}

\usepackage{graphicx}

\begin{document}

\newcommand\etal{\mbox{\textit{et al.}}}
\newcommand\Real{\mbox{Re}} 
\newcommand\Imag{\mbox{Im}} 
\newcommand\Rey{\mbox{\textit{Re}}} 
\newcommand\Pra{\mbox{\textit{Pr}}} 
\newcommand\Pec{\mbox{\textit{Pe}}} 

\title{Characteristics of the wavelength of ripples on icicles}

\author{K. Ueno}

\affiliation
{Graduate School of Engineering, Nagoya University, 
Chikusa, Nagoya 464-8603, Japan}

\email{k.ueno@kyudai.jp}


\begin{abstract}
It is known that the wavelength of the ripples on icicles in nature is of centimeter-scale. Such study on morphological instability of ice-water interface during ice growth from flowing supercooled water film with one side being a free surface has recently been made [K. Ueno, Phys. Rev. E \textbf{68}, 021603 (2003)]. This is a first theoretical study taking into account the influence of the shape of the water-air surface on the growth condition of infinitesimal disturbances of the ice-water interface. A simpler formula to determine the wavelength of the ripples than that in the previous paper is derived. It seems that the wavelength of ripples is insensitive to the water supply rates, diameters of the icicles and surrounding air temperatures. The details of dependence of the wavelengh of ripples on these parameters are investigated.
\end{abstract}

\maketitle

\section{introduction}
\begin{figure}[ht]
\begin{center}
\includegraphics[width=17cm,height=17cm,keepaspectratio,clip]{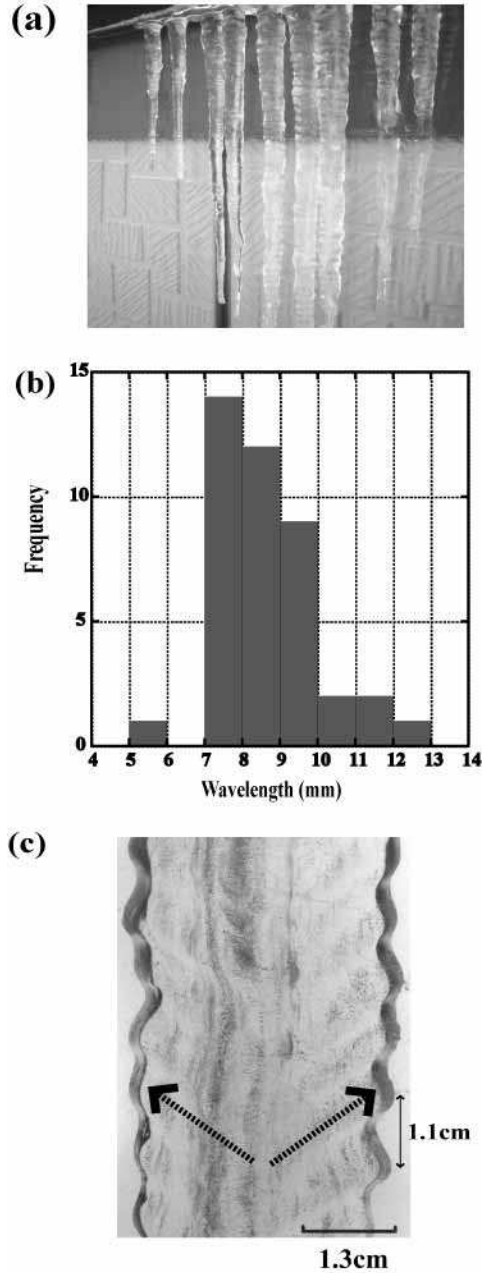}
\end{center}
\caption{(a) Icicles with ripples. 
(b) Frequency of the wavelengths of ripples on natural icicles.
(c) Vertical cross section of an iciccle with ripples.}
\label{fig:fig1}
\end{figure}

\begin{figure}[ht]
\begin{center}
\includegraphics[width=12cm,height=12cm,keepaspectratio,clip]{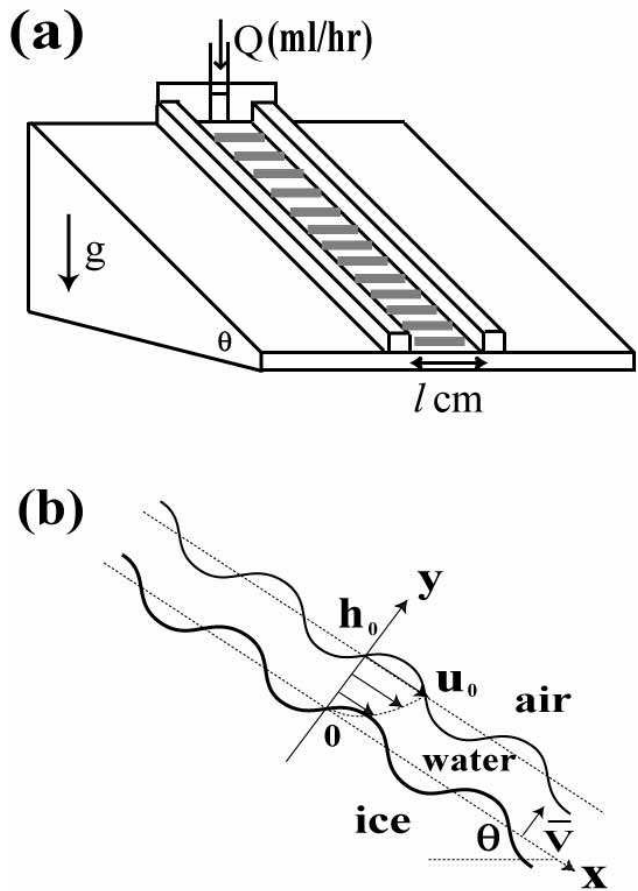}
\end{center}
\caption{(a) Schematic view of ice growth from a thin layer of supercooled water flowing down an inclined plane to the horizontal by angle $\theta$. $Q$ ml/hr is the water supply rate from the top and $l$ cm is the width of the gutter. Shaded portions within the gutter show nearly periodic ripples of ice. (b) Schematic view of vertical cross section $(x,y)$ of the inclined plane. The thick solid line is the ice-water interface and the solid line is the water-air surface. $h_{0}$ and $u_{0}$ are the mean thickness and the surface velocity of flowing liquid film in unperturbed state.}
\label{fig:fig2}
\end{figure}

Thin liquid films are ubiquitous entities in a variety of settings and display interesting dynamics. \cite{Oron97} The flow of liquid in a thin film down a solid wall is often observed in everyday life, for example, when rain water runs in a sheet down a window pane. 
The stability of a laminar flow of a viscous liquid running down an inclined plane was examined by Benjamin. \cite{Benjamin57} In this case, one side of the liquid film is a free surface and the other side is a rigid plane. 
However, when the liquid flows down on a solid-liquid interface accompanying a phase transition, the solid-liquid interface may not remain flat if a morphological instability occurs. Such an example is ring-like ripples on icicles as shown in Fig. \ref{fig:fig1}(a), which appears when icicles are covered with a thin film of flowing water surrounding by a cold air below $0^{\circ}$C. \cite{Maeno94}
Figure \ref{fig:fig1}(b) shows the frequency of the wavelengths of ripples on natural icicles with different diameters. This is based on the 41 data collected from photographs which were taken in many winters at various sites in Hokkaido in Japan. The growth conditions such as air temperature, water supply rate and wind speed, etc. are all different. Majorities of wavelengths are within $7 \sim 10$ mm and the mean wavelength is 8.5 mm.
In addition to the ripple formation, as shown by bold dashed lines in Fig. \ref{fig:fig1}(c), it is observed that many tiny air bubbles trapped in just upstream region of any protruded part of an icicle move in the upward direction during ice growth. \cite{Maeno94}
This indicates that an initial flat solid-liquid (ice-water) interface not only becomes unstable but also moves upwards. 
The similar pattern to the ripples on icicles can be experimentally produced on an inclined plane set in a cold room below $0^{\circ}$C by continuously supplying water from the top of the plane as schematically shown in Fig. \ref{fig:fig2}(a). \cite{Matsuda97} This experiment showed that the wavelength of ripple increases with a decrease in the angle $\theta$. At $\theta=\pi/2$, which nearly corresponds to the configuration of icicles, the average wavelength of the ripples was about 7.8 mm at the water supply rate $Q$=160 ml/hr. 

The fundamental building block of the morphological instability is the Mullins-Sekerka (MS) instability of a solidification front, which gives conditions for the growth of infinitesimal disturbances of a solid-liquid interface. \cite{Mullins63}
 According to the MS theory for a pure material, as a result of competition between destabilization due to the thermal diffusion and stabilization due to the Gibbs-Thomson (GT) effect, which is the melting temperature depression due to the curvature of the interface, a pattern with a specific wavelength of $\lambda_{\rm max}^{\rm MS}=2\pi\sqrt{3l_{d}d_{0}}$ on the order of microns is developed. \cite{Langer80,Caroli92}
The $\lambda_{\rm max}^{\rm MS}$ contains two characteristic lengths. One is the thermal diffusion length $l_{d}=\kappa_{l}/\bar{V}$ which is the thickness of an accumulation layer of the latent heat released by solidification. The layer is formed ahead of the solidification front in the semi-infinite liquid. This is usually macroscopic length determined by the values of the mean crystal growth velocity $\bar{V}$ and the thermal diffusivity $\kappa_{l}$ of the liquid. The other is the capillary length $d_{0}=T_{m}\Gamma C_{pl}/L^{2}$ associated with the solid-liquid interface tension $\Gamma$, where $T_{m}$ is the melting temperature, $C_{pl}$ is the specific heat at constant pressure of the liquid and $L$ is the latent heat of solidification per unit volume. This is a microscopic length of the order of angstroms.
Later, Cristini and Lowengrub (CL) revisited the linear thoery of the quasi-steady diffusional evolution of growing crystals in 3-dimension, developed by Mullins and Sekerka. \cite{Lowengrub02,Lowengrub04} They reposed the problem in terms of a critical flux rather than the critical radius in the MS theory. As a result, they revealed that the MS instability, that arises in the presence of constant undercooling, can be suppressed by maintaining flux conditions close to critical. This near-crtical conditions can be achieved by appropriately varying the undercooling in time.
In the MS and CL theories, the region of liquid is assumed to be semi-infinite. 
However, since the typical thickness of flowing water film on the surface of icicles is about 100 $\mu$m, the thermal diffusion layer cannot be formed in the liquid. Furthermore, the GT effect can be neglected because the wavelength of ripples on icicles is about 1 cm. Therefore, we have to consider different characteristic lengths from those in the MS theory and to develop quite a new mechanism for the ripple formation on icicles.

It is well known that the coupling of fluid flow in the liquid with phase transformation at the solid-liquid interface can result in the creation of new frontal instabilities. \cite{Davis01}
However, the morphological instability of solidification front in crystal growth from thin flowing liquid with one side being a free surface have never been examined. The problem involves both a solid-liquid interface and a liquid-air surface, in particular, both ice-water and water-air interfaces, as shown in Fig. \ref{fig:fig2}(b). No quantitative theory for predicting the wavelength of ripples has been proposed until a recent theoretical attempt to explain ripple formation on the surface of icicles is made. \cite{Ogawa02,Schewe02} According to their results of stability analysis of the solid-liquid interface, the instability of the solid-liquid interface occurs by the Laplace instability due to the thermal diffusion into the air, and the instability is suppressed by the effect of liquid flow in the film layer. They interpreted the suppression effect as the GT like effect, which makes the temperature distribution in the thin water layer uniform. They concluded that ripples of centimeter-scale wavelengths appears as a result of competition between these two effects. However, their dependence of the wavelength of the ripples on the angle of the inclined plane was deviated from the experimental result by Matsuda. Furthermore, their prediction that the ripples move in the downward direction was opposite to the observation mentioned above. Recently, a completely different mechanism for the ripple formation has been proposed. \cite{Ueno03,Ueno04} Here we briefly summarize some of results obtained from our theory.

In the first paper, \cite{Ueno03} the morphological instability of the solid-liquid interface of icicles was investigated by a linear stability analysis. We clarified that when the solid-liquid  interface and the liquid-air surface are coupled even if the modes are sinusoidal in the linear stability analysis, it needs to take into account the influence of the shape of the liquid-air surface of the liquid film on the growth condition of infinitesimal disturbances of the solid-liquid interface. It was found that restoring forces due to gravity and surface tension, although which act on the liquid-air surface, is an important factor for stabilization of morphological instability of the solid-liquid interface under appropriate thermodyanamic boundary conditions. We determined the wavelength of ripples on the incliend plane by varying the angle $\theta$ at the same water supply rate per width as the experiment by Matsuda. Our theoretical result showed that the wavelength increases with a decrease in the angle, which was in good agreement with experimental result (see FIG.4 in Ref. \onlinecite{Ueno03}). Moreover, our theory predicted that the solid-liquid interface moves in the upward direction, which is consistent with the observation in Fig. \ref{fig:fig1}(c). 

In the second paper, \cite{Ueno04} the magnitude of restoring forces due to gravity and surface tension changes depending on the wavenumber of the disturbed solid-liquid interface. As a result, the shape of the liquid-air surface is changed. There is a phase difference $\phi$ between a disturbed solid-liquid interface and a distribution of heat flux at the solid-liquid interface, whose magnitude is the cause for destabilization and/or stabilization of the solid-liquid interface. We showed that if $-\pi/2<\phi<0$, a disturbed solid-liquid interface is unstable to form ripples, while if $-\pi<\phi<-\pi/2$, a disturbed interface is stable and becomes flat. We also showed that the sign of $\phi$ is related to the direction of movement of the solid-liquid interface. Since $\phi<0$, the solid-liquid interface moves in the upward direction.

Our dispersion relation contains the mean thickness of the liquid film, 
$h_{0}=[2\nu/(g \sin\theta)]^{1/3}[3Q/(2l)]^{1/3}$,
where $Q$ is the water supply rate, $l$ is the span width of the gutter (see Fig. \ref{fig:fig2}(a)), $\nu$ is the kinematic viscosity of the liquid and $g$ is the gravitational acceleration. The experiment by Matsuda was performed for various temperatures of the cold room and the water supply rates under fixed span width $l=3$ cm. After trial and error, he found that ripples were clearly formed at the room temperature of $-8 \pm 1^{\circ}$C and the supply rate $Q=160$ ml/hr. \cite{Matsuda97} Hence, in the previous papers, \cite{Ueno03, Ueno04} we determined the wavelength of the ripples by varying the angle $\theta$ at the same water supply rate per width, $Q/l=160/3$ [(ml/hr)/cm], as the experiment by Matsuda. However, the equation of $h_{0}$ above indicates that the thickness also changes by varying the value of $Q/l$. Therefore we can expect the wavelength of the ripples to depend on $Q/l$. Nevertheless, it was claimed that the wavelength of the ripples on natural icicles is independent of  water supply rate and diameter of the icicles because the amount of $Q$ is expected to vary with the radius $R$ of icicles to keep the ratio $Q/l=Q/(2\pi R)$ during icicle growth. \cite{Ogawa02} In order to clarify it, we investigate the dependence of the wavelength on $Q/l$ and ambient temperature. We will present a simple formula to determine the wavelength of ripples. We also predict the dependence of the wavelength of ripples on the surface tension of the liquid-air and the dependence of the movement velocity of the solid-liquid interface on $Q/l$.

\section{Theoretical framework}

We give the theoretical framework developed in the previous papers \cite{Ueno03, Ueno04} with some modification and supplement. As schematically shown in Fig. \ref{fig:fig2}(a), the water at above nearly $0^{\circ}$C is supplied from the top nozzle at the rate $Q$ ml/hr and flowing down on an inclined plane with span width $l$ cm. \cite{Matsuda97} Since the apparatus is set in a cold room below $0^{\circ}$ C, the water becomes in the supercooled state as it flows down the plane. Ice grows from a portion of the supercooled water layer through which the latent heat of solidification is released into the ambient air below $0^{\circ}$C. 
The rest of the water drips from the lower edge of the plane. Our analysis is restricted to two dimensions $(x,y)$ of the vertical plane, and we assume that the region of the crystal is semi-infinite as shown in Fig. \ref{fig:fig2}(b) because the thickness of the liquid film is very thin compared to the typical redii of icicles. The $x$ axis is parallel to the inclined plane and the $y$ axis is normal to it. The position of $y=0$ is the solid-liquid interface in the frame of reference moving at the solid-liquid interface with an unperturbed solidification velocity $\bar{V}$. 

The velocity profile in the liquid film in the unperturbed state is given by the parabolic shear flow: \cite{Benjamin57, Landau59}
\begin{equation}
\bar{U}(y)=u_{0}\left\{2\frac{y}{h_{0}}-\left(\frac{y}{h_{0}}\right)^{2}\right\},
\label{eq:intro1}
\end{equation}
which is parallel to the $x$ axis as shown in Fig. \ref{fig:fig2}(b), where $h_{0}$ is the mean thickness of liquid film and $u_{0}$ is the surface velocity of liquid film in the unperturbed state. Then the water supply rate per width, $Q/l$, is given by
\begin{equation}
Q/l=u_{0}\int_{0}^{h_{0}}\left(2\frac{y}{h_{0}}-\frac{y^{2}}{h_{0}^{2}}\right)dy=\frac{2}{3}u_{0}h_{0}.
\label{eq:intro2}
\end{equation}
Since $u_{0}=gh_{0}^{2}\sin\theta/(2\nu)$, \cite{Landau59} $h_{0}$ and $u_{0}$ are not independent quantities. Therefore, it is convenient to express them by  experimentally controllable parameters $Q/l$ and $\theta$:
\begin{equation}
h_{0}=\left(\frac{2\nu}{g\sin\theta}\right)^{1/3}\left(\frac{3Q}{2l}\right)^{1/3}, 
\hspace{1cm}
u_{0}=\left(\frac{g\sin\theta}{2\nu}\right)^{1/3}\left(\frac{3Q}{2l}\right)^{2/3}.
\label{eq:intro3}
\end{equation}

\begin{figure}[ht]
\begin{center}
\includegraphics[width=8cm,height=8cm,keepaspectratio,clip]{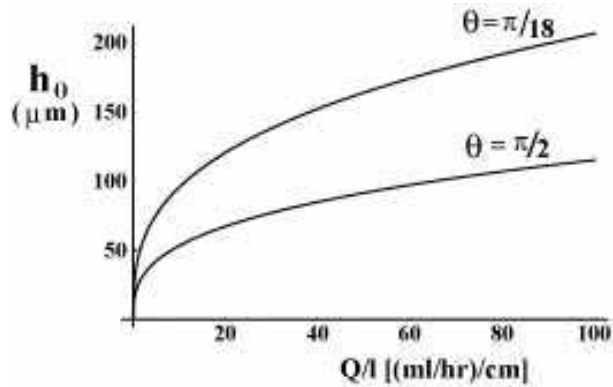}
\end{center}
\caption{Dependence of the thickness $h_{0}$ of the liquid film on the rate of volumetic flow per width, $Q/l$ [(ml/hr)/cm]  for $\theta=\pi/2$ and $\pi/18$.}
\label{fig:fig3}
\end{figure}

The following material properties of water are used: the kinematic viscosity, $\nu=1.8\times10^{-6}$ ${\rm m^{2}/s}$; the density, $\rho_{l}=1.0\times10^{3}$ ${\rm kg/m^{3}}$; the thermal diffusivity, $\kappa_{l}=1.3\times10^{-7}$ ${\rm m^{2}/s}$ and the specific heat at constant pressure, $C_{pl}=4.2\times 10^{6}$ J/(${\rm m}^{3}$$^{\circ}$C). The values of the surface tension of the water-air surface, the latent heat of solidification per unit volume and the ratio of the thermal conducutivity $K_{s}$ of ice to the thermal conducutivity $K_{l}$ of water are $\gamma=7.6\times10^{-2}$ N/m, $L=3.3 \times 10^{8}$ ${\rm J/m^{3}}$ and $n=K_{s}/K_{l}$=3.92, respectively.
Artificial icicles were made in a cold  laboratory room by controlling temperatures, water supply rates and wind speeds. \cite{Maeno94} The water supply rates were varied in the range of $Q=14.4 \sim 108$ ml/hr. Then the corresponding dripping rates from the tip of icicles were 15 $\sim$ 2 seconds. On the other hand, the water supply rate in the experiment by Matsuda \cite{Matsuda97} was kept at only $Q=160$ ml/hr. In the latter case, the values of $h_{0}$ and $u_{0}$ at $\theta=\pi/2$ and $l=3$ cm are 93 ${\rm \mu m}$ and 2.4 cm/s, respectively. Figure \ref{fig:fig3} shows the dependence of $h_{0}$ on $Q/l$ for $\theta=\pi/2$ and $\pi/18$. $h_{0}$ increases only gradually with $Q/l$. 
As dimensionless numbers, we define the Reynolds number, $\Rey\equiv u_{0}h_{0}/\nu=[3/(2\nu)](Q/l)$; the P${\rm \acute{e}}$clet number, $\Pec\equiv u_{0}h_{0}/\kappa_{l}=[3/(2\kappa_{l})](Q/l)$; the Froude number, $F\equiv u_{0}/(gh_{0})^{1/2}=\{\sin\theta/(2\nu)\}^{1/2}\{3Q/(2l)\}^{1/2}$ and the Weber number, $W\equiv \gamma/(\rho_{l}h_{0}u_{0}^{2})=\gamma /[\rho_{l}\{g\sin\theta/(2\nu)\}^{1/3}\{3Q/(2l)\}^{5/3}]$. We note that $\Rey$ and $\Pec$ depend on only $Q/l$, while $F$ and $W$ depend on both $Q/l$ and $\theta$. For the same value $Q/l=$160/3 [(ml/hr)/cm] as used in the experiment by Matsuda and $\theta=\pi/2$, these dimensionless numbers take the following values, $\Rey\sim1.23$, $\Pec\sim17.1$, $F\sim0.79$ and $W\sim1.42\times10^{3}$. 

\subsection{Governing equations}

The equations for the temperature in the liquid $T_{l}$, solid $T_{s}$ and air $T_{a}$ are given by
\begin{equation}
\frac{\partial T_{l}}{\partial t}+u\frac{\partial T_{l}}{\partial x}
+v\frac{\partial T_{l}}{\partial y}
=\kappa_{l}\left(\frac{\partial^{2} T_{l}}{\partial x^{2}}+\frac{\partial^{2} T_{l}}{\partial y^{2}}\right),
\label{eq:g2}
\end{equation}

\begin{equation}
\frac{\partial T_{s}}{\partial t}-\bar{V}\frac{\partial T_{s}}{\partial y}
=\kappa_{s}\left(\frac{\partial^{2} T_{s}}{\partial x^{2}}
+\frac{\partial^{2} T_{s}}{\partial y^{2}}\right),
\label{eq:g3}
\end{equation}

\begin{equation}
\frac{\partial T_{a}}{\partial t}-\bar{V}\frac{\partial T_{a}}{\partial y}
=\kappa_{a}\left(\frac{\partial^{2} T_{a}}{\partial x^{2}}
+\frac{\partial^{2} T_{a}}{\partial y^{2}}\right),
\label{eq:g4}
\end{equation}
where $t$ is time, $\kappa_{l}$, $\kappa_{s}$ and $\kappa_{a}$ are the thermal diffusivities of the liquid, solid and air, respectively. In Ref. \onlinecite{Ueno03}, $u$ and $v$ were the velocity components in the $x$ and $y$ direction measured in the laboratory frame, here we replace them with the velocity components measured in the frame of reference moving at the solid-liquid interface with $\bar{V}$. An experimental observation shows that the increase in the wind speed in the air leads to the increase in $\bar{V}$. \cite{Maeno94} We will show that the change in $\bar{V}$ affects the magnitude of amplification rate for disturbances of the solid-liquid interface, but the characteristic wavelength of the ripples does not change. As far as we are concerned with determining the wavelength, it is sufficient to assume no wind in the air.

The Navier-Stokes equations for $u$ and $v$ are
\begin{equation}
\frac{\partial u}{\partial t}+u\frac{\partial u}{\partial x}+v\frac{\partial u}{\partial y} 
=-\frac{1}{\rho_{l}}\frac{\partial p}{\partial x}
+\nu\left(\frac{\partial^{2}u}{\partial x^{2}}+\frac{\partial^{2}u}{\partial y^{2}}\right)+g\sin\theta,
\label{eq:g5} 
\end{equation}

\begin{equation}
\frac{\partial v}{\partial t}+u\frac{\partial v}{\partial x}+v\frac{\partial v}{\partial y}
=-\frac{1}{\rho_{l}}\frac{\partial p}{\partial y}
+\nu\left(\frac{\partial^{2}v}{\partial x^{2}}+\frac{\partial^{2}v}{\partial y^{2}}\right)-g\cos\theta, 
\label{eq:g6}
\end{equation}
where $p$ is the pressure, $\rho_{l}$ is the liquid density.
The equation of continuity is 
\begin{equation}
\frac{\partial u}{\partial x}+\frac{\partial v}{\partial y}=0.
\label{eq:g7}
\end{equation}
From (\ref{eq:g7}), $u$ and $v$ can be expressed as  
$u=u_{0}\partial \psi/\partial y$
and
$v=-u_{0}\partial \psi/\partial x$,
where $\psi$ is the stream function.

\subsection{Boundary conditions}

\subsubsection{Thermodynamic boundary conditions}

The continuity of the temperature at a disturbed solid-liquid interface $y=\zeta(t,x)$ is
\begin{equation}
T_{l}|_{y=\zeta}=T_{s}|_{y=\zeta}.
\label{eq:Tb1}
\end{equation}
The heat conservation at the solid-liquid interface is
\begin{equation}
L\left(\bar{V}+\frac{\partial \zeta}{\partial t} \right)
=K_{s}\frac{\partial T_{s}}{\partial y}\Big|_{y=\zeta}
      -K_{l}\frac{\partial T_{l}}{\partial y}\Big|_{y=\zeta},
\label{eq:Tb2}
\end{equation}
where $L$ is the latent heat per unit volume, and $K_{s}$ and $K_{l}$ are the thermal conductivities of the solid and liquid, respectively. 
On the other hand, the continuity of the temperature at a disturbed liquid-air surface $y=\xi(t,x)$ is
\begin{equation}
T_{l}|_{y=\xi}=T_{a}|_{y=\xi}=T_{la},
\label{eq:Tb3}
\end{equation}
where $T_{la}$ is a temperature at the liquid-air surface, which will be determined below.
The heat conservation at the liquid-air surface is
\begin{equation}
-K_{l}\frac{\partial T_{l}}{\partial y}\Big|_{y=\xi}
=-K_{a}\frac{\partial T_{a}}{\partial y}\Big|_{y=\xi},
\label{eq:Tb4}
\end{equation}
where $K_{a}$ is the thermal conductivity of the air.
We refer to these as the thermodynamic boundary conditions.

\subsubsection{Hydrodynamic boundary conditions}

We neglect the density difference between ice and water from the first.
Then both velocity components at the solid-liquid interface must satisfy
\begin{equation}
u|_{y=\zeta}=0,
\hspace{1cm}
v|_{y=\zeta}=0.
\label{eq:Hb1}
\end{equation}
The kinematic condition at the liquid-air surface is
\begin{equation}
\frac{\partial \xi}{\partial t}+u|_{y=\xi}\frac{\partial \xi}{\partial x}=v|_{y=\xi}.
\label{eq:Hb2}
\end{equation}
At the liquid-air surface the shear stress must vanish:
\begin{equation}
\frac{\partial u}{\partial y}\Big|_{y=\xi}
+\frac{\partial v}{\partial x}\Big|_{y=\xi}=0,
\label{eq:Hb3}
\end{equation}
and the normal stress including the stress induced by surface tension $\gamma$ of the liquid-air surface must just balance the pressure $P_{0}$ of the atmosphere:
\begin{equation}
-p|_{y=\xi}+2\rho_{l}\nu\frac{\partial v}{\partial y}\Big|_{y=\xi}
-\gamma\frac{\partial^{2}\xi}{\partial x^2}=-P_{0}.
\label{eq:Hb4}
\end{equation}
We refer to these as the hydrodynamic boundary conditions.

\subsection{Perturbations}

As seen in Fig. \ref{fig:fig1}(a), ring-like structure encircles the icicles and there is no noticeable azimuthal variation on the surface of the icicles. Therefore, it is sufficient to consider only a one dimensional perturbation in the $x$ direction of the solid-liquid interface:
\begin{equation}
\zeta(t,x)=\zeta_{k}\exp[\sigma t+i kx],
\label{eq:p6}
\end{equation}
where $k$ is the wavenumber and $\sigma=\sigma_{r}+i \sigma_{i}$, with $\sigma_{r}$ being the amplification rate and $v_{p} \equiv -\sigma_{i}/k$ being the phase velocity of the perturbation, and $\zeta_{k}$ is a small amplitude of the solid-liquid interface with the wavenumber $k$. 
We separate $\xi$, $T_{l}$, $T_{s}$, $T_{a}$, $\psi$ and $p$ into unperturbed steady fields and perturbed fields with prime as follows:
$\xi=h_{0}+\xi'$,
$T_{l}=\bar{T}_{l}+T'_{l}$,
$T_{s}=\bar{T}_{s}+T'_{s}$,
$T_{a}=\bar{T}_{a}+T'_{a}$,
$\psi=\bar{\psi}+\psi'$
and
$p=\bar{P}+p'$.
We suppose that the perturbations of the liquid-air surface, the temperature in the liquid, solid and air, the stream function, and pressure are expressed in the following forms:
\begin{equation}
\left(
\begin{array}{c}
\xi'(t,x) \\ T'_{l}(t,x,y) \\ T'_{s}(t,x,y) \\ T'_{a}(t,x,y) \\ \psi'(t,x,y) \\ p'(t,x,y) 
\end{array}
\right)
=
\left(
\begin{array}{c}
\xi_{k}  \\ g_{l}(y) \\ g_{s}(y) \\ g_{a}(y) \\ F(y) \\ \Pi(y) 
\end{array}
\right)
\exp[\sigma t+i kx],
\label{eq:pertset}
\end{equation}
where $\xi_{k}$, $g_{l}$, $g_{s}$, $g_{a}$, $F$ and $\Pi$ are the amplitudes of respective perturbations and they are assumed to be of the order of $\zeta_{k}$. 

The following calculations are based on a linear stability analysis taking into account only the first order of $\zeta_{k}$. 
Furthermore, we make two approximations. First, we use the long wavelength approximation, \cite{Oron97, Benjamin57} which is valid when the fluid film thickness is much less than the characteristic scale of ripples. Then we define a dimensionless wavenumber $\mu=kh_{0}$ and neglect higher orders in $\mu$. Second, we use the quasi-stationary approximation. \cite{Langer80, Caroli92} We can neglect the time dependence of perturbed part of temperature in each phase and that of liquid flow because these fields respond relatively rapidly to the slow development of the perturbation of solid-liquid interface. 

\begin{figure}[ht]
\begin{center}
\includegraphics[width=17cm,height=17cm,keepaspectratio,clip]{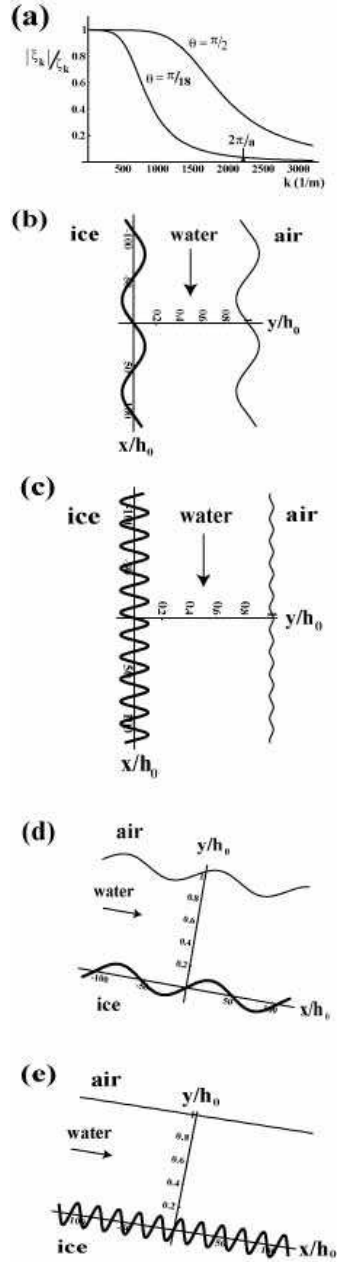}
\end{center}
\caption{(a) Dependence of the amplitude ratio $|\xi_{k}|/\zeta_{k}$ on the wavenumber $k$ for $\theta=\pi/2$ and $\pi/18$ at $Q/l=$160/3 [(ml/hr)/cm]. A typical value of the capillary length $a$ of water is about 2.8 mm, which corresponds to the value $k=2244$ ${\rm m^{-1}}$ indicated by the arrow. The shape of the water-air surface (solid lines) for a disturbed ice-water interface (thick solid lines) with a amplitude $\zeta_{k}/h_{0}=0.1$ when the wavelength (b) 1 cm ($k \sim 628$ ${\rm m}^{-1}$) and (c) 2 mm ($k \sim 3142$ ${\rm m}^{-1}$) for $\theta=\pi/2$, (d) 1 cm and (e) 2 mm for $\theta=\pi/18$. In the case of (b) and (c), $h_{0}\approx 93$ $\mu$m, while for (d) and (e), $h_{0}\approx 168$ $\mu$m.}
\label{fig:fig4}
\end{figure}

\subsection{Equation for the amplitude of perturbed part of the stream function and its solution}

When we substitute $\psi'$ and $p'$ into the equation for the perturbed parts of (\ref{eq:g5}) and (\ref{eq:g6}) and $\Pi$ is eliminated from them by cross differentiation, we obtain the Orr-Sommerfeld equation for the amplitude $F(y_{*})=f(y_{*})\zeta_{k}$ of the perturbed part of the stream function:
\begin{equation}
\frac{d^{4}f}{dy_{*}^{4}} =  i \mu \Rey\left\{(2y_{*}-y_{*}^{2})\frac{d^{2}f}{dy_{*}^{2}}+2f \right\},
\label{eq:solF1}
\end{equation}
where $y_{*}=y/h_{0}$. 
Since the values of $\mu\Rey$ is very small, we neglect the term on the right hand side of (\ref{eq:solF1}).  
Using the following linearized forms of hydrodynamic boundary conditions (\ref{eq:Hb1})-(\ref{eq:Hb4}), 
\begin{equation}
\frac{df}{dy^{*}}\Big|_{y_{*}=0}=-2, 
\hspace{1cm}
f|_{y_{*}=0}=0, 
\label{eq:LHB1}
\end{equation}
\begin{equation}
f|_{y_{*}=1}\zeta_{k}=-\xi_{k}, 
\label{eq:LHB2}
\end{equation}
\begin{equation}
\frac{d^{2}f}{dy_{*}^{2}}\Big|_{y_{*}=1}\zeta_{k}=2\xi_{k},
\label{eq:LHB3}
\end{equation}
\begin{equation}
\frac{d^{3}f}{dy_{*}^{3}}\Big|_{y_{*}=1}\zeta_{k}-i\alpha\xi_{k}=0, 
\label{eq:LHB4}
\end{equation}
the solution of $d^{4}f/dy_{*}^{4}=0$  is given by
\begin{equation}
f(y_{*})=
-2y_{*}+\frac{3(2-i\alpha)}{6-i\alpha}y_{*}^{2}+\frac{i\alpha}{6-i\alpha}y_{*}^{3}, 
\label{eq:solF4}
\end{equation}
where
\begin{equation}
\alpha=\frac{\mu \Rey \cos\theta}{F^{2}}+\mu^{3}\Rey W
      =2 \cot\theta h_{0}k+\frac{2}{\sin\theta}a^{2}h_{0}k^{3}
\label{eq:resforce}
\end{equation}
represents the restoring forces due to gravity and surface tension, \cite{Benjamin57} and $a=[\gamma/(\rho_{l}g)]^{1/2}$ is the capillary length associated with the surface tension $\gamma$ of the liquid-air surface. \cite{Landau59} Since the value of $W$ is very large as estimated above, the $\mu^{3}$ term in $\alpha$ cannot be neglected.

It should be noted that the difference between the stability analysis of a laminar flow and that of a solid-liquid interface. In the stability analysis of the laminar flow of a viscous liquid flowing down an inclined rigid plane, \cite{Benjamin57} the boundary conditions for the flat plane, $u|_{y=0}=0$ and $v|_{y=0}=0$, are used  instead of (\ref{eq:Hb1}). These and (\ref{eq:Hb2}) and (\ref{eq:Hb3}) are used to determine four constants arising from the solution of the fourth order differential equation (\ref{eq:solF1}) without neglecting the term of time derivative of $f$. The last boundary condition (\ref{eq:Hb4}) then be used to find the dispersion relation for the disturbance of the free surface. In the present case, however, there are five boundary conditions in (\ref{eq:Hb1})-(\ref{eq:Hb4}).  From (\ref{eq:Hb2}), the relation between the amplitude $\xi_{k}$ of the liquid-air surface and the amplitude $\zeta_{k}$ of the solid-liquid interface is obtained as follows
\begin{equation}
\xi_{k}=-f|_{y_{*}=1}\zeta_{k} 
       = \frac{6}{6-i\alpha}\zeta_{k}.
\label{eq:solF7}
\end{equation}
Figure \ref{fig:fig4}(a) shows the dependence of the amplitude ratio $|\xi_{k}|/\zeta_{k}$ 
on the wavenumber $k$ at $Q/l=160/3$ [(ml/hr)/cm] for $\theta=\pi/2$ and $\pi/18$.
$|\xi_{k}|/\zeta_{k}$ decreases monotonically with $k$ as shown in Fig. \ref{fig:fig4}(a). 

In the low wavenumber region in Fig. \ref{fig:fig4}(a) , the restoring forces due to gravity and surface tension on the liquid-air surface can be neglected because the value of $\alpha$ is very small, therefore, the amplitude of the liquid-air surface is almost the same as that of the solid-liquid interface as shown in Figs. \ref{fig:fig4}(b) and (d). In this case, fluid flows down along the disturbed solid-liquid interface. On the other hand, in the high wavenumber region in Fig. \ref{fig:fig4}(a), since the value of $\alpha$ is large, the shape of the liquid-air surface tends to be flat against the solid-liquid interface as shown in Figs. \ref{fig:fig4}(c) and (e). We can see that fluid flows down without feeling the shape of the solid-liquid interface. It should be noted that in the Ogawa and Furukawa's theory \cite{Ogawa02} the amplitude of the liquid-air surface is always the same as that of the solid-liquid interface irrespective of the wavenumber of the solid-liquid interface because the effect of the restoring forces on the liquid-air surface is not taken into account. 

\subsection{Equation for the amplitude of perturbed part of the temperature in the liquid and its solution}

Solving the equation for the unperturbed part of (\ref{eq:g2}) with the boundary conditions $\bar{T}_{l}=T_{m}$ at $y=0$ and $\bar{T}_{l}=T_{la}$ at $y=h_{0}$, the solution is 
$\bar{T}_{l}(y)=T_{m}-\bar{G}_{l}y$,
where $\bar{G}_{l} \equiv -d\bar{T}_{l}/dy|_{y=0}=(T_{m}-T_{la})/h_{0}$ is the unperturbed part of temperature gradient in the liquid film and $\bar{G}_{l}>0$ means that water is in supercooled state.  
When we substitute the assumed forms of $T'_{l}$ and $\psi'$ into the equation for the perturbed part of (\ref{eq:g2}), we obtain the equation for $g_{l}$:
\begin{equation}
\frac{d^{2}g_{l}}{dz^{2}}-i\mu \Pec (1-z^{2})g_{l} 
=i\mu \Pec f(z)\bar{G}_{l}\zeta_{k},
\label{eq:gsol1}
\end{equation}
where we have introduced a new varaible, $z=1-y/h_{0}$.

When putting the right hand side of (\ref{eq:gsol1}) equal to zero, the independent solutions are given by 
\begin{equation}
\phi_{1}(z)=\exp\left(-\frac{1}{2}(-i\mu \Pec)^{1/2}z^2\right)\,\!_{1}F_{1}\left(\frac{1}{4}\left[1+\frac{i\mu \Pec}{(-i\mu \Pec)^{1/2}}\right],\frac{1}{2},(-i\mu \Pec)^{1/2}z^2 \right),
\label{eq:gsol2}
\end{equation}
\begin{equation}
\phi_{2}(z)=z\exp\left(-\frac{1}{2}(-i\mu \Pec)^{1/2}z^2\right)\,\!_{1}F_{1}\left(\frac{1}{2}+\frac{1}{4}\left[1+\frac{i\mu \Pec}{(-i\mu \Pec)^{1/2}}\right],\frac{3}{2},(-i\mu \Pec)^{1/2}z^2 \right),
\label{eq:gsol3}
\end{equation}
where $_{1}F_{1}$ is the confluent hypergeometric function. \cite{Jeffreys56}
Noting that the Wronskian of the two solutions $\phi_{1}(z)$ and $\phi_{2}(z)$ equals to unity, the general solution of (\ref{eq:gsol1}) can be written as
\begin{equation}
g_{l}(z)
=B_{1}\phi_{1}(z)+B_{2}\phi_{2}(z) 
+i\mu \Pec \bar{G}_{l}\zeta_{k}
\int_{0}^{z}\left\{\phi_{2}(z)\phi_{1}(z')-\phi_{1}(z)\phi_{2}(z')\right\}f(z')dz',
\label{eq:gsol4}
\end{equation}
where $B_{1}$ and $B_{2}$ are unknown constants to be determined from the thermodynamic boundary conditions at the liquid-air surface.

With the boundary conditions $\bar{T}_{a}=T_{la}$ at $y=h_{0}$ and $\bar{T}_{a}=T_{\infty}$ at $y=h_{0}+\delta$, the solution of the equation for the unperturbed part of (\ref{eq:g4}) is
\begin{equation}
\bar{T}_{a}(y)=T_{la}-\bar{G}_{a}(y-h_{0}),
\label{eq:gsol5}
\end{equation}
where $\bar{G}_{a} \equiv -d\bar{T}_{a}/dy|_{y=h_{0}}=(T_{la}-T_{\infty})/\delta$ is the unperturbed part of the temperature gradient in the air at the liquid-air surface, and we regard $\delta$ as the thickness of a thermal boundary layer. Strictly speaking, the thermal boundary layer, which is created around icicles warmer than surrounding air, can be provided from similarity solutions for the coupled Navier-Stokes and heat transport equations in the Boussinesq approximation. However, we assume that the solution  (\ref{eq:gsol5}) is different from the exact form of the similarity solution only by the multiplication of an order one constant. \cite{Short06} Therefore, we estimate the value of  $\delta$ from Eq. (4) in Ref. \onlinecite{Short06}.

Substituting the assumed form of $T'_{a}$ into the equation for the perturbed part of (\ref{eq:g4}), we obtain
\begin{equation}
g_{a}(y)=T_{ka} \exp[-k(y-h_{0})],
\label{eq:gsol6}
\end{equation}
where $T_{ka}$ is the amplitude of perturbed part of the temperature in the air. 
Linearizing (\ref{eq:Tb3}) at $y=h_{0}$ gives to the zeroth order in $\xi_{k}$, 
\begin{equation}
\bar{T}_{l}|_{y=h_{0}}=\bar{T}_{a}|_{y=h_{0}}=T_{la},
\label{eq:gsol7}
\end{equation}
and to the first order in $\xi_{k}$, noting that $g_{l}|_{z=0}=B_{1}$ since $\phi_{1}|_{z=0}=1$ and $\phi_{2}|_{z=0}=0$, 
\begin{equation}
-\bar{G}_{l}\xi_{k}+B_{1}=-\bar{G}_{a}\xi_{k}+T_{ka}=0.
\label{eq:gsol8}
\end{equation}
Linearizing (\ref{eq:Tb4}) at $y=h_{0}$ gives to the zeroth order in $\xi_{k}$,
\begin{equation}
K_{l}\bar{G}_{l}=K_{a}\bar{G}_{a},
\label{eq:gsol9}
\end{equation}
and to the first order in $\xi_{k}$, noting that $dg_{l}/dz|_{z=0}=B_{2}$ since $d\phi_{1}/dz|_{z=0}=0$ and $d\phi_{2}/dz|_{z=0}=1$,
\begin{equation}
K_{l}B_{2}=\mu K_{a}T_{ka}.
\label{eq:gsol10}
\end{equation}
From (\ref{eq:gsol8})-(\ref{eq:gsol10}), $B_{1}$ and $B_{2}$ are determined and (\ref{eq:gsol4}) becomes
\begin{equation}
g_{l}(z)
=\left[-f|_{z=0}\left\{\phi_{1}(z)+\mu\phi_{2}(z)\right\} 
+i \mu \Pec
\int_{0}^{z}\left\{\phi_{2}(z)\phi_{1}(z')-\phi_{1}(z)\phi_{2}(z')\right\}f(z')dz'\right]\bar{G}_{l}\zeta_{k} 
\equiv  H_{l}(z)\bar{G}_{l}\zeta_{k},
\label{eq:sol11}
\end{equation}
where we have used the relation $\xi_{k}=-f|_{z=0}\zeta_{k}$. 
Hence, (\ref{eq:gsol1}) can be written as 
\begin{equation}
\frac{d^{2}H_{l}(z)}{dz^{2}}-i\mu \Pec (1-z^{2})H_{l}(z) 
=i\mu \Pec f(z).
\label{eq:gsol12}
\end{equation}
It is found that the perturbed part of heat transport in the liquid film in the long wavelength region is dominated by mean shear flow $\bar{U}$ of the second term on the left hand side and normal flow $v'$ generated from a disturbance of the solid-liquid interface on the right hand side. The heat transport by thermal diffusion can be neglected because this term appears from the second order in $\mu$. 

\subsection{Dispersion relation}
Linearizing (\ref{eq:Tb2}) at $y=0$ gives to the first order in $\zeta_{k}$, we obtain
\begin{equation}
L\sigma \zeta_{k}=
K_{s}\frac{dg_{s}}{dy}\Big|_{y=0}
-K_{l}\frac{dg_{l}}{dy}\Big|_{y=0}.
\label{eq:dis2}
\end{equation}
With the condition that the disturbance in the solid must vanish far from the solid-liquid interface, 
the solution of the equation for the perturbed part of (\ref{eq:g3}) is given by
\begin{equation}
g_{s}(y)=T_{ks}\exp(ky),
\label{eq:dis3}
\end{equation}
where $T_{ks}$ is the amplitude of perturbed part of the temperature of the solid, which is determined from the perturbed part of (\ref{eq:Tb1}) as follows
\begin{equation}
-\bar{G}_{l}\zeta_{k}+g_{l}|_{y=0}
=T_{ks}.
\label{eq:dis4}
\end{equation}

Substituting  (\ref{eq:sol11}) and (\ref{eq:dis3}) into (\ref{eq:dis2}), the dispersion relation for the disturbance of the solid-liquid interface is obtained:
\begin{equation}
\sigma = \frac{\bar{V}}{h_{0}}\left\{\frac{dH_{l}}{dz}\Big|_{z=1}
          +n\mu\left(H_{l}|_{z=1}-1\right)\right\},
\label{eq:sol(i)10}
\end{equation}
where $n=K_{s}/K_{l}$, and
\begin{equation}
H_{l}|_{z=1}
=-f|_{z=0}\left\{\phi_{1}|_{z=1}+\mu\phi_{2}|_{z=1}\right\} 
+i\mu \Pec
\int_{0}^{1}\left\{\phi_{2}|_{z=1}\phi_{1}(z')-\phi_{1}|_{z=1}\phi_{2}(z')\right\}f(z')dz', 
\label{eq:sol(i)8}
\end{equation}
and
\begin{equation}
\frac{dH_{l}}{dz}\Big|_{z=1}
=-f|_{z=0}\left\{\frac{d\phi_{1}}{dz}\Big|_{z=1}
+\mu\frac{d\phi_{2}}{dz}\Big|_{z=1}\right\} 
+i \mu \Pec
\int_{0}^{1}\left\{\frac{d\phi_{2}}{dz}\Big|_{z=1}\phi_{1}(z')
-\frac{d\phi_{1}}{dz}\Big|_{z=1}\phi_{2}(z')\right\}f(z')dz'.
\label{eq:sol(i)9}
\end{equation}
We can express  (\ref{eq:gsol2}) and  (\ref{eq:gsol3}) by the expansion in terms of $\mu\Pec$ as follows
\begin{equation}
\phi_{1}(z) 
=1+i\left(\frac{1}{2}z^{2}-\frac{1}{12}z^{4}\right)\mu\Pec 
+\left(-\frac{1}{24}z^{4}+\frac{7}{360}z^{6}-\frac{1}{672}z^{8}\right)(\mu \Pec)^2
+\cdots, 
\label{eq:phi1}
\end{equation}
\begin{equation}
\phi_{2}(z) 
=z+i\left(\frac{1}{6}z^{3}-\frac{1}{20}z^{5}\right)\mu \Pec 
+\left(-\frac{1}{120}z^{5}+\frac{13}{2520}z^{7}-\frac{1}{1440}z^{9}\right)(\mu \Pec)^{2}
+\cdots .
\label{eq:phi2}
\end{equation}
Putting $z=1$ in (\ref{eq:phi1}) and (\ref{eq:phi2}), we obtain
$\phi_{1}|_{z=1}=1+5i(\mu\Pec)/12-239(\mu\Pec)^{2}/10080+\cdots$
and
$\phi_{2}|_{z=1}=1+7i(\mu\Pec)/60-13(\mu\Pec)^{2}/3360+\cdots$. 
The ratio of the term of second order in $\mu\Pec$ to the first one in $\phi_{1}|_{z=1}$ and $\phi_{2}|_{z=1}$ are about $5.7\times 10^{-2}\mu\Pec$ and $3.3\times 10^{-2}\mu\Pec$,  respectively. When the value of $\mu\Pec=[3^{4/3}/(2\kappa_{l})][\nu/(g\sin\theta)]^{1/3}(Q/l)^{4/3}(2\pi/\lambda)$ takes about 100,  the first term and second one become the same order. This is possible when $Q/l \sim 1000$ [(ml/hr)/cm] for typical wavelength $\lambda$ of ripples on icicles.
For natural icicles with radii $R$, $Q/l$ does not necessarily take a definite value. Since the typical values of $Q$ over icicles are on the order of tens of ml/hr and their radii are usually in the range of $1\sim 10$ cm, \cite{Maeno94} we vary the value of $Q/l$ over the range of $10 \sim 100$ [(ml/hr)/cm]. 
As far as we are concerned in such a range of $Q/l$, $\mu\Pec\sim O(1)$ for the wavelength of ripples on icicles, then it is sufficient to consider up to the first term in $\mu\Pec$ of (\ref{eq:phi1}) and (\ref{eq:phi2}) in the following calculations. 

\section{Results}

The amplification rate $\sigma_{r}$ and the phase velocity $v_{p}=-\sigma_{i}/k$ are obtained from the real and imaginary part of (\ref{eq:sol(i)10}), respectively, as follows:
\begin{equation}
\sigma_{r}
=\frac{\bar{V}}{h_{0}}
\left[\frac{-\frac{3}{2}\alpha(\mu\Pec)
+\mu\left\{36-\frac{3}{2}\alpha(\mu\Pec)\right\}}{36+\alpha^{2}}
+n\mu
\frac{-\frac{7}{10}\alpha(\mu\Pec)-\alpha^{2}
+\mu\left\{36-\frac{7}{10}\alpha(\mu\Pec)\right\}}{36+\alpha^{2}}\right],
\label{eq:Ueno-sigma_r}
\end{equation}
\begin{equation}
v_{p}=-\frac{\bar{V}}{\mu}
\left[\frac{-\frac{1}{4}\alpha^{2}(\mu\Pec)
+\mu\left\{6\alpha+9(\mu\Pec)\right\}}{36+\alpha^{2}} 
+n\mu\frac{6\alpha-\frac{7}{60}\alpha^{2}(\mu\Pec)
+\mu\left\{6\alpha+\frac{21}{5}(\mu\Pec)\right\}}{36+\alpha^{2}}\right].
\label{eq:Ueno-Vp}
\end{equation}

With the assumption of $\bar{T}_{s}=T_{m}$ in the solid, linearizing (\ref{eq:Tb2}) at $y=0$ gives to the zeroth order in $\zeta_{k}$,
\begin{equation}
\bar{V}=\frac{C_{pl}\kappa_{l}(T_{m}-T_{la})}{Lh_{0}},
\label{eq:sol(i)1}
\end{equation}
where $C_{pl}$ is the specific heat at constant pressure of the liquid.
From (\ref{eq:sol(i)1}), the limit $\bar{V}h_{0}/\kappa_{l}\ll 1$ corresponds to
\begin{equation}
\frac{C_{pl}(T_{m}-T_{la})}{L} \ll 1.
\label{eq:sol(i)2}
\end{equation}
This indicates that the supercooling $T_{m}-T_{la}$ in the liquid is much smaller than $L/C_{pl}$. In this case, thermal diffusion layer of the thickness $\kappa_{l}/\bar{V}$ ahead of the solidification front in the liquid is not formed because $h_{0} \ll \kappa_{l}/\bar{V}$, and the latent heat is released into the air through the thin liquid film. For the typical values $\bar{V} \approx 10^{-6}$ m/s and $h_{0}\approx 10^{-4}$ m, this condition is indeed satisfied. 

\begin{figure}[ht]
\begin{center}
\includegraphics[width=10cm,height=10cm,keepaspectratio,clip]{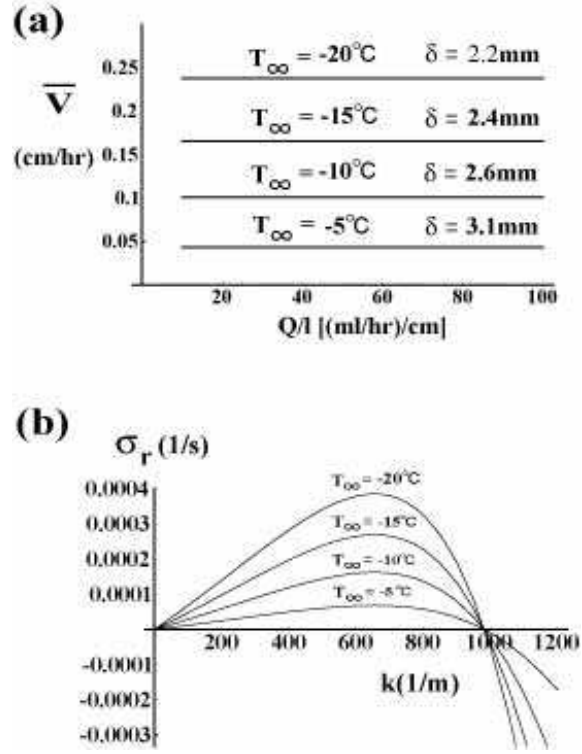}
\end{center}
\caption{(a) The dependence of $\bar{V}$ on $Q/l$ for various air temperatures $T_{\infty}$. $\delta$ is the thickness of the boundary layer around icicles. (b) Amplification rate $\sigma_{r}$ vs the wavenumber $k$ at $Q/l=50$ [(ml/hr)/cm] and $\theta=\pi/2$ for various $T_{\infty}$.}
\label{fig:fig5}
\end{figure}

From (\ref{eq:gsol9}), $T_{la}$ in (\ref{eq:Tb3}) is obtained as
\begin{equation}
T_{la}=\frac{T_{m}+\frac{K_{a}}{K_{l}}\frac{h_{0}}{\delta}T_{\infty}}
{1+\frac{K_{a}}{K_{l}}\frac{h_{0}}{\delta}}.
\label{eq:sol(i)3}
\end{equation}
If the values of $T_{\infty}$, $h_{0}$ and $\delta$ are given, the value of $T_{la}$ is determined.
Substituting (\ref{eq:sol(i)3}) into (\ref{eq:sol(i)1}) yields
\begin{equation}
\bar{V}=\frac{K_{l}}{Lh_{0}}\frac{T_{m}-T_{\infty}}{1+\frac{K_{l}}{K_{a}}\frac{\delta}{h_{0}}}.
\label{eq:sol(i)4}
\end{equation}
Since $K_{l}/K_{a}\approx 22.7 \gg 1$ and $\delta \gg h_{0}$, (\ref{eq:sol(i)4}) can be approximated as
\begin{equation}
\bar{V} \approx \frac{K_{a}}{L}\frac{T_{m}-T_{\infty}}{\delta}.
\label{eq:sol(i)5}
\end{equation}
Indeed, we estimate the value of $\bar{V}$  from (\ref{eq:sol(i)4}) for various $Q/l$ and $T_{\infty}$ using Eq. (4) in Ref. \onlinecite{Short06} when $C=1$ and $z=0.1$ m. The value of $\bar{V}$ increases with a decrease in the ambient air temperature $T_{\infty}$ as shown in Fig. \ref{fig:fig5}(a). However, it is found that $\bar{V}$ remains constant against $Q/l$, hence $\bar{V}$ does not depend on $h_{0}$ as indicated in (\ref{eq:sol(i)5}). This behavior agrees with an observation that growth velocity of diameter of icicles is almost constant with an increase in water supply rate (see Fig. 6b in Ref. \onlinecite{Maeno94}). 
 
Figure \ref{fig:fig5}(b) shows the amplification rate $\sigma_{r}$ for disturbances of the wavenumber $k$ of the solid-liquid interface. It is found that the change in $\bar{V}$ by ambient air temperature affects the magnitude of maximum point of $\sigma_{r}$, but does not make change the characteristic wavelength $\lambda_{\rm max}$ of the ripples, which is defined at the wavenumber at which $\sigma_{r}$ takes a maximum value. 
Heat transport through the air surrounding icicles can be greatly influenced by the presence of forced convection. Such an enhancement of heat transport around growing icicles increases the mean growth rate $\bar{V}$ and results in early appearence of ripples because the value of $1/\sigma_{r}$ becomes shorter without changing the wavelength of ripples. This theoretical prediction should be checked by experiment of icicle growth under both calm environment and forced convection. 

\begin{figure}[ht]
\begin{center}
\includegraphics[width=17cm,height=17cm,keepaspectratio,clip]{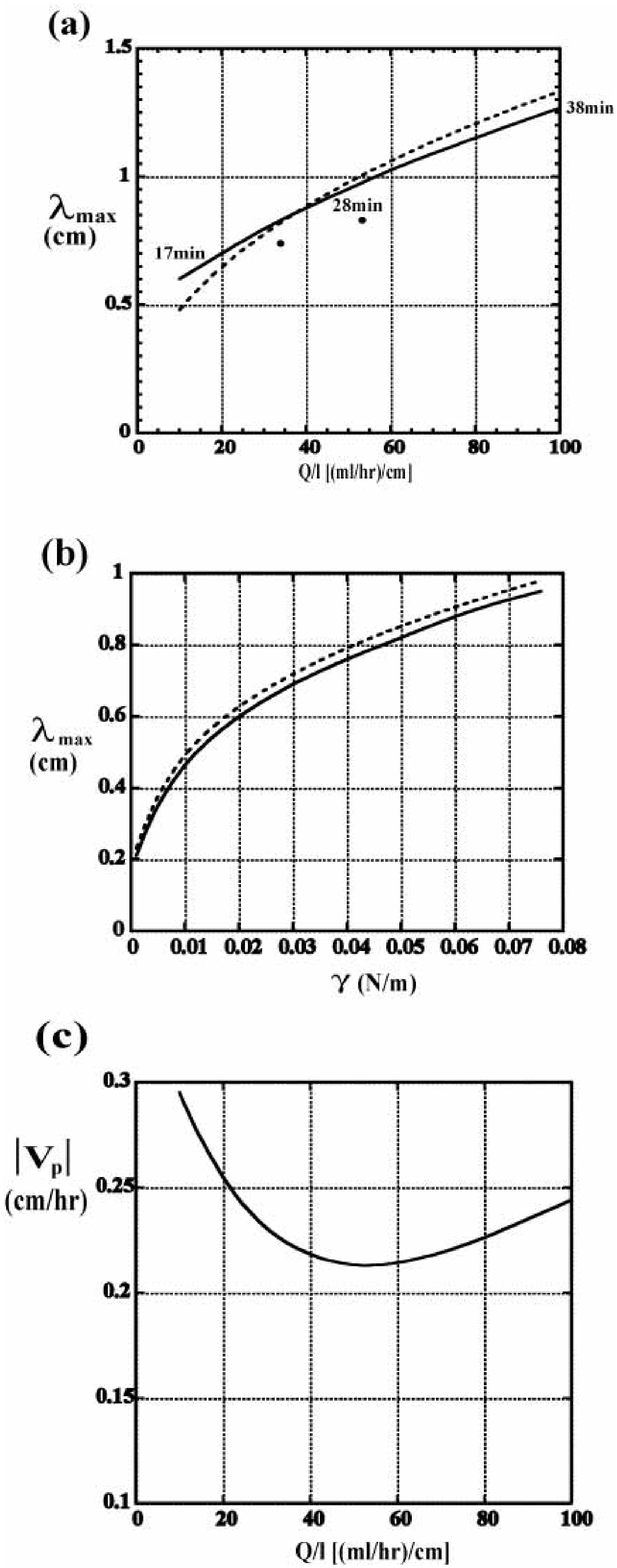}
\end{center}
\caption{(a) Dependence of $\lambda_{\rm max}$ on $Q/l$ [(ml/hr)/cm] at $\theta=\pi/2$. 
Thick solid line: $\lambda_{\rm max}$ obtained from (\ref{eq:Ueno-sigma_r}), 
thick dashed line: $\lambda_{\rm max}\approx 0.17\times (Q/l)^{4/9}$ (cm) 
obtained from the formula $\lambda_{\rm max}=2\pi(a^{2}h_{0}\Pec/3)^{1/3}$.
Two solid circles are experimental results by Matsuda. \cite{Matsuda97} 
(b) Dependence of $\lambda_{\rm max}$ on the surface tension $\gamma$ at $Q/l=50$ [(ml/hr)/cm] and $\theta=\pi/2$. 
Thick solid line: $\lambda_{\rm max}$ obtained from (\ref{eq:Ueno-sigma_r}),
thick dashed line: $\lambda_{\rm max}\approx 2.31\times \gamma^{1/3}$ (cm)
obtained from the formula $\lambda_{\rm max}=2\pi(a^{2}h_{0}\Pec/3)^{1/3}$.
(c) Dependence of $v_{p}$ on $Q/l$ at $\theta=\pi/2$ when $\bar{V}=10^{-6}$ m/s.}
\label{fig:fig6}
\end{figure}

Figure \ref{fig:fig6}(a) shows the dependence of $\lambda_{\rm max}$ on $Q/l$ in the range of $10 \sim 100$ [(ml/hr)/cm] at $\theta=\pi/2$. The thick solid line is the result obtained from (\ref{eq:Ueno-sigma_r}). Since $\mu\Pec\sim 1$ and $\alpha \sim 1$ for the wavelength of ripples on icicles, (\ref{eq:Ueno-sigma_r}) can be approximated as 
\begin{equation}
\sigma_{r}
\approx \frac{\bar{V}}{h_{0}}
\left[\frac{-\frac{3}{2}\alpha(\mu\Pec)+36\mu}{36}\right].
\label{eq:Ueno-approx-sigma_r}
\end{equation}
Here we have neglected the second term in (\ref{eq:Ueno-sigma_r}) because this term is smaller than the first one and the wavenumber at which $\sigma_{r}$ takes a maximum value is almost the same without the second term. However, we note that this formula holds near $\theta=\pi/2$ because the value of $\alpha$ increases with a decrease in $\theta$. 
From $d\sigma_{r}/dk=0$ in (\ref{eq:Ueno-approx-sigma_r}), we obtain a simple formula to determine the wavelength of ripples:
\begin{equation}
\lambda_{\rm max}=2\pi\left(\frac{a^{2}h_{0}\Pec}{3}\right)^{1/3}
\approx 0.17 \left(\frac{Q}{l}\right)^{4/9}{\rm (cm)}.
\end{equation}
This result is shown by the thick dashed line in Fig. \ref{fig:fig6}(a), which gives a good approximation to the thick solid line. $\lambda_{\rm max}$ increases only gradually with $Q/l$. Indeed, $\lambda_{\rm max}=5 \sim 13$ mm for $Q/l=10 \sim 100$ ml/hr. These wavelengths are within the distribution of the wavelength of ripples on natural icicles in Fig. \ref{fig:fig1}(b). The numbers 17 min, 28 min and 38 min attached near the thick solid line represent the characteristic time of appearence of ripples for $Q/l$=10, 50 and 100 [(ml/hr)/cm], respectively. These values have been estimated from the value of $1/\sigma_{r}$ at the most unstable mode when $\bar{V}=10^{-6}$ m/s. It is found that the characteristic time increases with an increase in $Q/l$. On the other hand, in the Ogawa and Furukawa's theory the formula to get the wavelength of ripples is given by $\lambda_{\rm max}\approx (2\pi/2.2)h_{0}\Pec \approx 2.3\times10^{-3}(Q/l)^{4/3}$ cm. \cite{Ogawa02} This leads to $\lambda_{\rm max}=0.5 \sim 11$ mm for $Q/l=10 \sim 100$ ml/hr. However, the two formulae are obviously different. Each $\lambda_{\rm max}$ shows completely different power law behavior with respect to $Q/l$. $\lambda_{\rm max}$ obtained from our theory contains two characteristic lengths, the capillary length $a=[\gamma/(\rho_{l}g)]^{1/2}$ and the mean thickness $h_{0}=[2\nu/(g\sin\theta)]^{1/3}[3Q/(2l)]^{1/3}$ of the liquid film, and one dimensionless number $\Pec=[3/(2\kappa_{l})](Q/l)$.
On the other hand, $\lambda_{\rm max}$ obtained from the Ogawa and Furukawa's  theory contains only one characteristic length $h_{0}$ and one dimensionless number $\Pec$. Two solid circles in Fig. \ref{fig:fig6}(a) are experimental result by Matsuda. \cite{Matsuda97} One is the mean wavelength of 7.4 mm for ripples formed on a wooden round stick with the diameter of 1.5 cm, and the other is the mean wavelength of 8.3 mm for ripples formed on a vertical plane within the gutter with the width 3 cm. In both experiments, water supply rate from the top was fixed at 160 ml/hr. There is no experimental data for other $Q/l$. Matsuda mentions that ice itself does not grow when the water supply rate is too small or too large.\cite{Matsuda97} The similar thing is also mentioned by Makkonen. \cite{Makkonen88} These suggest that there may be an appropriate range of $Q/l$ to clearly form ripples. The present theory cannot explain this fact because the growth rate $\bar{V}$ is constant even in the case of too small and too large water supply rate as shown in Fig. \ref{fig:fig5}(a). In order to confirm our theoretical prediction for the dependence of $\lambda_{\rm max}$ on $Q/l$, it needs to do experiments to see the behavior of the growth rate of ice for much water supply rate than that shown in Fig. 6b in Ref. \onlinecite{Maeno94} and to modify the present theory which takes into account the dependence of the mean growth rate of the radius of icicles on the water supply rate. 

In our theory, the capillary length associated with the surface tension of the liquid-air surface is one of the important characteristic lengths in determining the wavelength of ripples. Our theory predicts that the wavelength of ripples becomes shorter with the reduction of the surface tension as shown in Fig. \ref{fig:fig6}(b). The thick solid line is the result obtained from (\ref{eq:Ueno-sigma_r}). The thick dashed line obtained from the formula $\lambda_{\rm max}=2\pi(a^{2}h_{0}\Pec/3)^{1/3}$ gives a good approximation to the thick solid line. It needs to measure the wavelength of ripples when reducing the surface tension of water by adding a surfactant.

Figure \ref{fig:fig6}(c) shows the dependence of the phase velocity $|v_{p}|$ on $Q/l$ at $\theta=\pi/2$ and $\bar{V}=10^{-6}$ m/s. Here we define the value of $v_{p}$ at the wavenumber at the maximum point of $\sigma_{r}$. At this wavenumver, (\ref{eq:Ueno-Vp}) takes a negative value. 
It shows that the change in $v_{p}$ for various $Q/l$ is small, which takes almost the same order as the mean growth rate $\bar{V}=10^{-6}$ m/s=0.36 cm/hr. The negative sign means that the solid-liquid interface moves against the primary shear flow, which is consistent with the observation that many tiny air bubbles are trapped in just upstream region of any protruded part of the icicle and line up in the upward direction during icicle growth as shown by bold dashed lines in Fig. \ref{fig:fig1}(c). \cite{Maeno94} It needs to measure precisely the magnitude of velocity of the motion of the ripples under carefully control of $Q/l$ and air temperature.

\section{\label{sec:sumdis}Summary and Discussion}
We have investigated the morphological instability of the ice-water interface during ice growth from flowing supercooled water film with one side being a free surface. We have obtained a simple formula, $\lambda_{\rm max}=2\pi(a^2h_{0}\Pec/3)^{1/3}$, to determine the wavelength of ripples.  $\lambda_{\rm{max}}$ depends on the two characteristic lengths, the thickness $h_{0}$ of the water film and the capillary length $a$ associated with the surface tension of the water-air surface, and one dimensionless number $\Pec$, which is the ratio of heat transfer of liquid flow to that of thermal diffusion. It should be stressed that $\lambda_{\rm max}$ includes the fluid effect in the thin water film througth the parameter $Q/l$. This is contrast to $\lambda_{\rm max}^{\rm MS}=2\pi\sqrt{l_{d}d_{0}}$ obtained from the MS theory, where $l_{d}$ is the thermal diffusion length and $d_{0}$ is the capillary length asoociated with the solid-liquid interface tension. It is found that each characteristic length scale is completely different. Icicles in nature possess a wide range of $Q/l$. Nevertheless, the wavelength of the ripples on icicles seems to be insensitive to the supply rate $Q$ and the radius $R$ of icicles. This is not because $Q$ varies with $R$ to keep the ratio $Q/R$ as discussed in Ref. \onlinecite{Ogawa02}, but because $\lambda_{\rm{max}}$ does not strongly change for the range of typical values of $Q/R$ on icicles. Indeed, we have found that $\lambda_{\rm{max}}$ increases only gradually with $Q/l$. 

\begin{acknowledgements}
The author would like to thank N. Maeno for providing me with the data of Fig. 1(b) and a picture of Fig. 1(c) 
This work was supported by a Grant-in-Aid for the 21st Century COE ``Frontiers of Computational Science".
\end{acknowledgements}

\end{document}